\documentclass[journal]{IEEEtran}

\usepackage[T1]{fontenc}
\usepackage[utf8]{inputenc}
\usepackage{amsmath,amsfonts,amssymb}
\usepackage{stmaryrd} 
\usepackage{mathrsfs}
\usepackage{xspace}
\usepackage{listings}
\usepackage{newlfont}
\usepackage{url}
\usepackage{hyperref}
\usepackage{float}
\usepackage[table,usenames,dvipsnames]{xcolor}
\usepackage{graphicx}
\usepackage{widetext}
\usepackage{fancyhdr}
\usepackage{alltt}
\usepackage{epsfig,psfrag}
\usepackage{graphicx}
\usepackage{subcaption}
\usepackage{srcltx}
\usepackage{theorem}
\usepackage{latexsym}
\usepackage{textcomp}
\usepackage{setspace}
\usepackage{cite}
\usepackage{array}
\usepackage{mdwmath}
\usepackage{mdwtab}
\usepackage{enumerate}

\newcommand{\benum}{\begin{enumerate}}
\newcommand{\eenum}{\end{enumerate}}
\newcommand{\nc}{\newcommand}
\newcommand{\rnc}{\renewcommand}
\newcommand{\bvb}{\begin{verbatim}}
\newcommand{\evb}{\end{verbatim}}

\nc{\<}{\langle}
\rnc{\>}{\rangle}
\nc{\bq}{\textbf}
\nc{\m}{\textrm}
\nc{\bb}{\mathbb}
\nc{\til}{\texttildelow}
\nc{\dps}{\displaystyle}
\rnc{\l}{\left(}\rnc{\r}{\right)}
\nc{\lc}{\left\{}\nc{\rc}{\right\}}
\nc{\lb}{\left[}\nc{\rb}{\right]}
\nc{\ba}[1]{\begin{array}{#1}}
\nc{\ea}{\end{array}}       
\nc{\ra}{\rightarrow}
\nc{\li}{\left |}
\nc{\ri}{\right |}
\nc{\pde}[2]{\frac{\partial #1}{\partial #2}}
\nc{\ode}[2]{\frac{d #1}{d #2}}
\nc{\odee}[3]{\frac{d^{#3} #1}{d #2^{#3}}}
\nc{\pdee}[3]{\frac{\partial^{#3} #1}{\partial #2^{#3}}}
\nc{\bn}{\begin{enumerate}}
\nc{\en}{\end{enumerate}}
\nc{\bt}{\begin{theorem}}
\nc{\et}{\end{theorem}}
\nc{\y}[1]{\lambda_{#1}}
\nc{\ninf}{{\oplus}^{-\infty}}
\nc{\pinf}{{\oplus}^{+\infty}}
\nc{\nninf}{{\otimes}^{-\infty}}
\nc{\ppinf}{{\otimes}^{+\infty}}
\nc{\ir}{\mathbb{I}\mathbb{R}}
\nc{\closure}[2][3]{{}\mkern#1mu\overline{\mkern-#1mu#2}} 
\nc{\ep}{\mathcal{E}_{P}}
\nc{\mr}{\mathcal{M}_{r}}
\nc{\mfa}{\mathcal{M}_{f,a}}
\nc{\mfp}{\mathcal{M}_{f,p}}
\nc{\mt}{\m{T}}
\nc{\dB}{\textit{dB}}

\nc{\VM}{\delta} 
\nc{\GM}{\Theta}
\nc{\PM}{\phi} 
\nc{\x}{\mathbf{x}}

\newtheorem{theorem}{Theorem}

\usepackage[commandnameprefix=ifneeded, commentmarkup=footnote]{changes}
\definechangesauthor[name={Alessandro Pinto}, color=red!40]{AP}
\definechangesauthor[name={Timothy Wang}, color=blue!40]{TW}

\begin{document}
\title{Survey of Human Models for Verification of Human-Machine Systems
} 


\author{Timothy E. Wang \thanks{Raytheon Technologies Research Center, East Hartford, CT, USA, wangte@rtx.com}
Alessandro Pinto \thanks{Raytheon Technologies Research Center, Berkeley, CA, USA, alessandro.pinto@rtx.com}%
}

\maketitle

\begin{abstract}
We survey the landscape of human operator modeling ranging from the early cognitive models 
developed in artificial intelligence to more recent 
formal task models developed for model-checking of human machine interactions.   
We review human performance modeling and human factors studies in
the context of aviation, and models of how the pilot interacts with automation in the cockpit. 
The purpose of the survey is to assess the applicability of available 
state-of-the-art models of the human operators for the design, verification and validation of future safety-critical
aviation systems that exhibit higher-level of autonomy, but still require human operators in the loop.  
These systems include the single-pilot aircraft and NextGen air traffic management. 
We discuss the gaps in existing models and propose future research to address them.
\end{abstract}

  

\section{Introduction} 
\label{sec:introduction}
The development of increasingly capable Artificial Intelligence (AI) and Machine Learning (ML) techniques, 
is inspiring both industry and academia to build systems that automate the execution of certain tasks in a variety of applications. 
These systems do not replace humans, but rather help to simplify operations and, at minimum, the human is still responsible to provide goals to the intelligent agents that execute these tasks. Moreover, humans are involved in being part of the execution process by providing information that automation cannot infer reliably, or by making decisions whose consequences cannot be accurately predicted by the autonomous systems. Thus, at least for the foreseeable future, autonomous systems will be required to interact with humans, collaborating and complementing each other towards delivering systems that can effectively operate in complex environments. The design of these human-machine systems (HMS) challenges the adequacy of current development processes especially when the application domain is safety critical, such as commercial aviation. 
These applications require a high-level of assurance and demand for evidence that the closed-loop system satisfies stringent safety and performance requirements. 

\begin{figure}
    \includegraphics[width=\columnwidth]{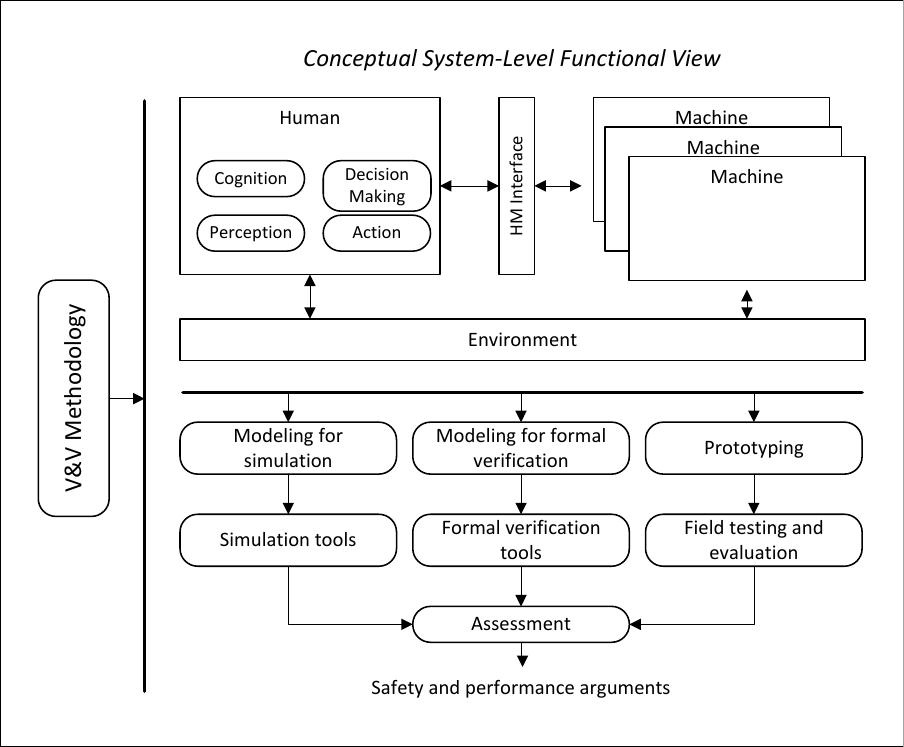}
    \caption{Typical framework for design, analysis and verification of human-machine systems from functional decomposition to safety arguments.} 
    \label{fig:mhms}
\end{figure}

Verification and Validation (V\&V) are critical steps in a high-assurance design process. A V\&V methodology (Figure \ref{fig:mhms}) defines the set of tools and techniques to arrive at a safety argument, including the functional view of the system, the modeling activities, and how to use their results. Several techniques can be used to generate evidence such as simulation, formal verification, and field-testing. While the latter is necessary at some point in the verification process, it involves prototyping and testing campaigns that are expensive. Field-testing alone becomes impractical for autonomous systems due to the large number of scenarios required to achieve coverage. For example, it is predicted that an autonomous car would need to drive in realistic conditions for billions of miles before reaching confidence on its safety statistics\cite{kalra_driving_2016}. Simulation and formal verification are therefore appealing alternatives aiming at reducing the amount of field-testing while at the same time providing confidence on safety and performance. These techniques rely on models that are machine-interpretable and that can be used in a systematic exploration of plausible scenarios to find errors. These models must include three sub-components (Figure \ref{fig:mhms}): the system comprising one or multiple machines, the environment, and the human. 
Developing these models is hard in general. Models of machines can be built by engineers, as they are the subject of the design activity. The environment and the human models are typically not readily available and require substantial development effort. Among these two, the environment model captures the physical space in which the system operates and can be developed by leveraging knowledge of the application domain. However, the development of capability, performance, and behavioral models of  humans is considerably harder, as it must account for behavioral, neurological, sociological, psychological, and systems aspects that are difficult to formalize. 

Given the importance of a virtual engineering platform for autonomy applications, and the complexity associated with building models of humans, this survey focuses on identifying the major currents and gaps in the ocean of knowledge about human modeling for the specification, design, verification and validation of HMS. We survey both simulation and formal models for the major functions of a human decision cycle including perception, cognition, decision-making, and action, and we also provide extensive background on performance and error models.

\subsection{Overview of Human Models Surveyed} 

The human models surveyed in this paper span all 3 of these types: \emph{descriptive}, \emph{computational}, and \emph{formal}. 
Descriptive models represent the quality of the human behavior without producing any quantitative outputs. 
Computational models, in contrast, can be simulated or executed to produce quantitative predictions about the 
performance of the HMS. 
Formal models are rigorous mathematical descriptions that can be used for formal analysis.
Formal analysis differs from computational analysis in the sense that the former generates proofs that guarantee that certain properties of the system are satisfied under all modeled environment, while the latter can be used to test the system in a given scenario or sets of scenarios.  Formal methods are used in analysis, but they can also be used for synthesizing systems with given property, thereby enabling correct-by-construction design.

In terms of specification language, descriptive models are typically given in a natural language such as English. 
Computational and formal models are specified using a rich variety
of mathematical and/or formal languages: 
static equations such as linear and nonlinear equations including neural networks, 
dynamical equations such as differential equations, Markov decision processes, Kalman filters, 
statistical models such as Bayesian nets, Structured Equation Models,
state-machines such as FSM, Petri-nets, programming languages as in cognitive architectures, and logic such as modal predicate logic. 


A typical property to be verified is whether the HMS satisfies certain performance or safety constraints. Oftentimes, performance models are constructed using certain hidden states or latent variables of a human such as boredom.  For that reason, we also survey the models that are based on those hidden states in the context of using them for design and verification. 

The survey starts in Section~\ref{sec:models-of-humans} by 
discussing some of the fundamental concepts of human modeling (see the human box in Figure~\ref{fig:mhms}) 
that are related to the OODA loop~\cite{Boy87}.  These ``first principles of human'', come from scientific studies of 
human behaviors and cognition from fields such as psychology, sociology and artificial intelligence.
This is followed by Section~\ref{sec:latent} with an introduction to various hidden states  modelled in the human factors field along with a discussion of the various ``failure modes'' of a human. 
This is motivated by our system-theoretic interpretation of the human, with inputs and outputs, 
unobservable hidden states, and faults.  
In Section~\ref{sec:formal-models}, we review formal models. 
The notion of an epistemic model is introduced and some of the challenges in
formally verifying epistemic models are discussed.  
In Section~\ref{sec:simulations} we review various computational or simulation models used in the analysis 
and verification of human-machine systems including integrated frameworks for the simulation of 
performance and behavior of a pilot in the cockpit. 
Finally, we conclude this paper with a set of recommendations on the research and development efforts needed to build a unified framework for modeling and V\&V of future human-autonomy systems. 
In summary, the models survey in this paper are categorized in Table~\ref{tab:summary0}. 
\begin{table}[htp]
    \caption{Summary of surveyed models}
\begin{tabular}{ |p{3cm}||p{4.5cm}| }
    \hline
	Category & Models Surveyed \\
	\hline
    Application & human-aviation systems \\
    \hline
	Capabilities  & visual and perception, cognition, decision making, actuation \\
    \hline
	Specification language & natural language, static equations,  dynamic equations, stochastic equations, logic, programs, state machines\\ 
    \hline
	Verification techniques & simulation, formal verification\\
    \hline
    Hidden states & workload, trust, situational awareness, boredom, complacency, vigilance \\  
	\hline
\end{tabular}
\label{tab:summary0}
\end{table}

\subsection{Related Work}

There are prior works in the literature which also survey 
human modeling for the purpose of verifying human-machine interaction. 

For example, in~\cite{bolton2013using}, Bolton et al. provides perhaps the earliest known surveys 
 on human models for formal verification of human-computer interaction (HCI).  
In addition, Weyers et al. in their book~\cite{weyers2017handbook} provide a comprehensive collection of works on 
 formal methods usage in human-machine interaction illustrated on a set of well-known 
case studies in the HMI domain, and a detailed overview of past, current and future developments of formal methods in HCI.  
Unlike those previous works, this paper surveys all models of human operators for
the verification of human-machine systems, not just models for formal verification, 
but also models for simulation and execution.  
The prior works also surveyed broader set of application domains 
unlike this work which is primarily driven by the need for a human-autonomy 
virtual engineering platform as shown in Figure~\ref{fig:mhms}.

\section{Foundational Concepts} 
\label{sec:models-of-humans}
This section reviews some foundational concepts of human behavior including   
perception, cognition, decision-making and actuation. Motivated by Boyd's OODA loop from his writing 
on combat operational process~\cite{Boy87}, 
the OODA loop consists of four stages: observe, orient, decide and act. 
\emph{Observe} is the ability of humans to perceive and sense the environment around them. In Section~\ref{subsec:visual}, some basic theories of human perception (primarily visual) are surveyed. \emph{Orient} is the ability of humans to process the information coming from their sensory sub-systems, understand it, and make predictions based on such understanding.  This ability is provided by the cognitive processes of the human. In Section~\ref{subsec:cognitive}, we discuss briefly the theories of human cognition. 
\emph{Decide} is the ability of humans to create course of actions based on goals, utility or payoff functions, and available information. Fundamental concepts in human decision-making are surveyed in Section~\ref{subsec:decide}.  Finally, \emph{act} is the execution of plans, which is briefly discussed in Section~\ref{subsec:act}. 

\subsection{Perception} 
\label{subsec:visual}
A fundamental framework for building models of human perception is Signal Detection Theory (SDT)~\cite{green1966signal}, which is is one of most successful quantitative theories of human performance~\cite{wickens2002elementary}. 
It is an experimental framework in which the human participants are asked to detect the 
signal within a noisy input stimulus.  The participant's reaction
time and accuracy are recorded, and are used to build a model of sensitivity and response bias. 

In human perception modeling, one of the central questions that researchers seek to answer is how
efficiently can a human operator find a specific area of information on a visual display, 
a task also known as
\emph{visual search}.  A noisy stimulus, in the context of visual search, can be the presence of
items of distractions on a display with the target item.  Examples of using the SDT framework
for building models of human visual search include~\cite{Ver01,Eck00}, and numerous other works which can
be found in the survey paper~\cite{Wol98}. 

A notable work in human visual attention is the feature integration theory of Treisman \&
Gelade~\cite{Tre80}, which attempted to answer the question about the serial and/or parallel characteristics of the visual process.  Their model hypothesizes that feature search is mostly parallel while
other visual processes are serial.  In contrast, the work of Bundeson in~\cite{Bun90} contains a model with a strictly parallel architecture.  The parallel argument can be traced back to skepticism about distinguishing the serial processes from the parallel process in works such as~\cite{Tow90}.
Some early theories on the mental effort of human visual attention and processing can be found in~\cite{kahneman1973attention} where the performance is dependent on a single pool of undifferentiated resources.  This work later progressed into theory of multiple-resources~\cite{Wic91} (MRT), where attention and processing resources are divided into certain differentiated pools.  


\subsection{Cognition}
\label{subsec:cognitive}
A theory of cognition describes human cognitive
processes such as reasoning, problem solving, decision-making, language processing, memory, and learning. 
The development of a theory of cognition is a multi-disciplinary effort that spans 
linguistics, artificial intelligence, psychology, and sociological, neurological and behavior sciences. 
A comprehensive review of cognition is beyond the scope of this paper.  
Instead, we survey the efforts to unify and codify cognitive principles into a common framework for 
building computational models of human cognition, reasoning and learning.  
These common frameworks, also referred to as cognitive architectures, 
are collections of structural and mechanistic descriptions 
of ``what human cognitive behaviors have in common"~\cite{lehman1996gentle}.  
Cognitive architecture models are the realization of the unified theories of 
human cognition and reasoning in the form of computer programs that 
simulate underlying cognitive processes of the human, and predict various cognitive outcomes or phenomena.  
They are based on both
fundamental theories about the human cognitive process and assumptions
about the domain being modelled such as language acquisition, helicopter pilot
training, expert behaviors, or implicit learning. 
One of the earliest cognitive models  
was the EPAM (Elementary Perceiver and Memorizer)~\cite{Sim64} by Simon
et al.  It is a quantitative model based on the verbal learning of nonsense
syllables, which predicts the rate of verbal learning as a function of
parameters such as familiarity, meaningfulness, and similarity. 
More recently, the trend has been to unify the theories of the cognitive processes, and then
formalize them as a basis for a programming language
environment which allows the end-users to write programs that are executable
models of the human mind.

\subsection{Decision Making} 
\label{subsec:decide}

Models of human decision making range from 
game theory in economics~\cite{Von07},  where each agent minimizes or maximizes a utility function,
to sub-optimal descriptive models (behavioral economics)~\cite{Tve92}, where the model tries to reproduce the
heuristics of actual human decisions based on various human idiosyncrasies such as bounded
rationality~\cite{Gig02} and over-confidence in their knowledge~\cite{Gig91}.  
Other works on informal heuristics of human decision making include~\cite{Gig96,Gig99,Pay93}.
In a similar vein, the notion of satisficing decision-making, which originated in Herbert Simon's 1947 
work on administrative theory and refined over the years in subsequent works such as~\cite{simon1955behavioral},
relaxes the notion of the maximization of utility to simplify decision making under large
uncertainty and poorly defined utility function~\cite{reverdy2017satisficing}.  
In the satisficing decision-making framework, the limited informational and 
computational capacities of the human are taken into account. 
A decision is made if a simple payoff function exceeds a certain threshold value, which captures the process of making decisions of unknown optimality that are ``good enough'' for a given problem.

From the field of psychology, the lens model from Brunswik's 1952 work~\cite{Bru52} 
became an important conceptual framework for modeling human decision-making.  
The main idea behind the lens model is that one can quantitatively capture 
a human judgment of a criterion using a finite set of cues or observations related to the criterion. 
Examples include a manager making a judgment on the future success of a candidate (the criterion) based on
the behavior that the candidate exhibits during an interview (the cues), or a financial analyst
making a prediction on the future movements of a stock (the criterion) based on current social media news (the cues). 
The typical lens model consists of two halves. Each half is a multi-linear regression model. The right half 
captures the relations between the cues and the human judgment.  
The left half captures the relations between the cues and the actual outcome of the criterion.  
A meta-survey of the lens model can be found in~\cite{Kar08}.

\subsection{Actuation} 
\label{subsec:act} 
Actuation refers to the motor outputs of the human operator. The typical
motor outputs that are modeled in human factors research include simple tasks
such as pointing, dragging, clicking, gazing, steering, etc. Other forms of
actuation such as verbal or non-verbal body movements are not described here.
A fundamental model of actuation is the Hyman-Hick law~\cite{Hym53,Hic52}, 
which hypothesized that the difficulty of a motor task is proportional to
the entropy of the situation. More commonly, the Hyman-Hick law is known in the 
context of decision-making i.e. 
more choices equates to higher entropy which 
means the person will take longer to make a decision.  

While Hyman-Hick established an 
information-theoretic foundation for the performance modeling of motor tasks, it 
did not become the dominant modeling
framework. Instead the most ubiquitous modeling framework in quantifying human
motor performance is Fitts' law~\cite{Fit54,Mac92}, which relates the difficulty of a
task to the amount of information that needs to be transmitted over a
hypothetical communication channel of limited capacity. The time performance of the motor task is 
linearly correlated with the amount of information transmitted by the task.  
Consider the pointing task in which a person is required to move a
finger to a target on the screen.  The signal for this task is the distance $D$ to
the target, and the noise is the width $W$ of the target. A model capturing the difficulty of 
the task would be the entropy of the task i.e. $\log{\frac{D}{W}}$.  
Fitt's model has been generalized to other tasks beyond
finger pointing such as  mouse, trackball, stylus, touchpad, helmet-mounted sights, eye tracker and joy stick~\cite{Sou04}, and
also generalized to predict errors in motor tasks~\cite{Wob08}.  

Beyond Fitt's law, a more recent work by Accot et al.~\cite{Accms} extended the
informational-theoretic framework to cover complex motor tasks that are more
trajectory based.  Examples of these motor tasks include steering, drawing
curves, navigating through nested menus, moving in a 3D world, which becomes
more relevant as interfaces move beyond the simple point and click paradigm.


\section{Hidden States and Failure Modes} 
\label{sec:latent}

This section introduces various hidden states or latent variables used in models
for predicting human performance in HMS.
Many of these hidden states such as workload, situational awareness, trust, vigilance, and boredom,  
are understood qualitatively but are difficult to define precisely and to measure directly. 
Despite their controversial status~\cite{dekker2004human}, they have proven to be useful for creating 
engineering models for the design and V\&V of HMS.  

In addition to performance, the V\&V process must guarantee safety of a HMS.  
Traditionally, the human operator 
is not taken into account in safety analysis except as a mitigation factor for the system hazards identified.  
However, humans fail too. Taking humans into account in the hazard analysis phase of the design process requires modeling their failure modes. For this reason, we introduce models of human errors, and specifically mode confusion.   

\subsection{Workload} 
\label{subsec:workload}
Mental workload is extensively studied in human factors~\cite{moray2013mental}.  It stems from the study of attention and information processing by cognitive psychologists~\cite{navon1979economy}.  Mental workload is conceived to be a multi-dimensional construct of time stress, mental effort and psychological stress~\cite{sheridan1979toward,gopher1986workload,reid1988subjective} that is independent of behavior and performance~\cite{yeh1988dissociation}.  There is no commonly accepted formal definition of workload, however generally speaking, workload is often defined in terms of a relation between the mental resources required by a task and the available mental resources that are supplied by the human.  In~\cite{navon1979economy}, it is noted that every human has a bounded set of resources available at any given time.  Early workload literature hypothesized that this pool of resources was singular and undifferentiated~\cite{gopher1984psychophysics}. The prevailing current theory of workload is that the pool of resources is multi-dimensional i.e. that it is divided into different channels of capacities~\cite{wickens1980processing,jex1988measuring,Wic91}. The idea of different channels of capacities of resources was introduced in works such as~\cite{kahneman1973attention} and eventually developed into what is known as the multiple resource theory (MRT)~\cite{wickens1991processing}.  In MRT, the human operator has separate resource blocks that span multiple dimensions of input modalities, cognitive stages, and code processing.  In the modalities dimension, we find the visual, auditory, tactile and olfactory blocks. In the stages dimension, we find the perception, processing, and action blocks.  Some empirical evidence providing the separation between perception, processing and action resources can be found in~\cite{isreal1980p300}. In the codes dimension, there are the symbolic (spatial) and the linguistic (verbal) blocks.  The codes dimension is based on the hemispheric laterality framework where some tasks are exclusively executed by a certain hemisphere of the brain (the left brain versus right brain paradigm).  In MRT, the degradation of the human performance is simply the result of overloading in any of the individual resource blocks due to the performance of one or more tasks. Using the MRT model, the phenomenon of cross-task interference can be measured by quantifying the amount and the type of resources taken up by each of the tasks. 

Other workload theories in the literature can be interpreted as a special instance of the multiple resource theory either by using a hemispheric dimension in~\cite{polson1988task,friedman1981hemispheres}, 
by expanding one dimension into other sub-dimensions in~\cite{boles1998simultaneous}, 
or by restricting to only one dimension in~\cite{pashler1994dual}.  
The response selection bottleneck theory of Pashler~\cite{pashler1994dual} is an example of 
multiple resource theory that captures concurrent task interference only within the stages dimension.  
In Pashler's model, the interference between the two tasks is total, meaning that the response selection portion (processing stage) 
of the task acts like a semaphore which uses all of the available resources of the human operator. 
Underlying the bottleneck theory is the theory of psychological refractory period (PRP)~\cite{smith1967theories} which establishes that the human operator's ability to respond to external stimuli is completely 
degraded during a certain period right after it has received an earlier stimulus. 
The PRP phenomenon is captured in the perceptual modality within the EPIC cognitive architecture~\cite{Mey97}. 

Workload is difficult to measure as it is context and operator-dependent.  Disassociations can occur frequently between the different measures of workload~\cite{yeh1988dissociation}. Workload varies based on many factors such as individual's capabilities,  environment factors, and the difficulty, type, timing, and criticality of the task~\cite{meshkati1988toward}. There are 3 predominant basis to measure workload: performance, subjective evaluation ratings, and psycho-physiological measurements. Inferring workload using task performance measurements can be problematic. There is empirical evidence that low workload leads to degraded performance~\cite{young2002malleable}.  The latter can be caused by, for example, the induced boredom in the human operator, which has an effect on human vigilance~\cite{sawin1995effects}. 
It has been noted that more often than not, task performance fails to reflect the resource re-allocation prompted by the changes in the task demands~\cite{gopher1986workload}.  For example, the human operator frequently adjusts its effort in response to more challenging tasks. 
Workload measures can disassociate from performance measures~\cite{yeh1988dissociation} for both simple or challenging tasks.  Nonetheless performance-based measures of either primary or secondary tasks is a widely-used approach to evaluate workload and human-machine system performance in various contexts~\cite{eggemeier1991performance,hart1990workload}.

\subsection{Situational Awareness}
\label{subsec:sa} 

Situation awareness (SA) is defined by Endsley in~\cite{End00} as ``knowing what is going on around you'' with an inherent notion of knowing what is important. 
Many other definitions of SA have been developed either in the domain of
aviation or more general applications.  A good survey of SA definitions
relating to aircraft piloting can be found in~\cite{Dom94}.  

Endsley gave a detailed descriptive model of SA in~\cite{End95} as having a hierarchical structure, with the first level being perception, the second level being  comprehension, and the third level being projection.  
The last level captures the capability of the human operator
to not only being able to comprehend the current state of the situation, but also to understand what is going to happen in the future.
In the Endsley model, SA is a separate stage from decision making
because it represents the construction of the human operator's internal model
of the external world.  Any decisions that the operator made in constructing
this internal world has no impact on the external world. Moreover it has been
observed that human operators can make poor decisions even during a state of
excellent situational awareness.

\subsection{Vigilance, Boredom, and Complacency} 
\label{subsec:folk}

The notion of complacency is mentioned frequently in the aviation human factors literature~\cite{billings1984human} and the term has long been absorbed into the vernacular of aviation accident reports~\cite{ntsb_aar121,funk1999flight} as a causal factor.  In a survey of leading airline captains~\cite{wiener1981complacency}, more than half thought that complacency was the leading cause of aviation accidents. The notion is considered by some as controversial.  It is criticized, along with other concepts such as SA, mental workload, and trust  as ``folk'' models in~\cite{dekker2004human} with little underlying scientific basis and more common sense appeal. A counter-argument is given in~\cite{parasuraman2008situation} supporting these notions with an extensive data-base of empirical evidences, computational models and psychological foundations supporting SA, mental workload and trust.

Wiener in~\cite{wiener1981complacency} discussed complacency as a low level of suspicion but suggested more empirical research to understand any underlying mechanisms behind complacency.  In~\cite{parasuraman1993performance}, complacency is a term used to explain the higher rate of mis-detections of automation failure when the reliability of automation is increased.  While there is no consensus on the exact definition of complacency, it is suggested in~\cite{parasuraman2010complacency} that a working definition could be derived from the features of complacency common to both aviation accidental analyses and human performance studies.  This set of features include human monitoring of automation, a sub-optimal frequency of monitoring of the automation~\cite{moray2000attention}, and finally observable degradation in system performance due to the sub-optimal monitoring.  A conceptual model of complacency is given~\cite{parasuraman2010complacency}.  In this model, complacency is a dynamic variable that varies overtime through interactions with the automation, and affects the attention allocation of the operator, which in turns may or may not lead to errors in the operation of machines.

Vigilance can be interpreted as the lack of complacency~\cite{parasuraman1993performance}. The notion of vigilance is usually defined
in the human factors literature as the ability of the human operator to focus attention on a signal source 
for extended period of time while remaining alert to a particular signal~\cite{hancock2017nature}.  
The study of vigilance started with Mackworth's work in World War 2 to understand why sonar operators miss certain weak signals indicating the presence of submarines at the end of their watch.  
A review of recent research on vigilance in~\cite{warm2008vigilance} shows, through various subjective reports and more objective physiological measure of blood flow within the brain, that maintaining vigilance require high workload, and that a decrease in vigilance over extended period of time can be explained by the attentional resource theory of MRT.  

As discussed in the survey in~\cite{cummings2016boredom}, boredom is closely related to vigilance, complacency and 
attention management.  The prevailing view is that boredom is 
caused by the monotony or repetitiveness of tasks that require 
constant attention~\cite{smith1981boredom,prinzel1997task}. Alternatively, a more recent view is that low task loading such as passive monitoring of autonomous systems~\cite{cummings2016boredom} also contributes to boredom. Given the current trend of increasing autonomy in aviation, boredom leading to complacency~\cite{wiener1988human} could result in increasing number of problems. 

\subsection{Trust}
\label{subsec:folk2}
Trust is often defined in the various human sciences as some expectation by the human of certain outcome behavior.  As stated by the Harvard sociologist Bernard Barber in~\cite{barber1983logic}, it is the ``expectation of technically competent role performance.''  Some of the earliest works on human trust in the automation domain can be found in~\cite{sheridan1984research,muir1987trust} where trust was raised as a factor in the context of modeling the supervisory control of computer or automation.
In these early works, the main hypothesis is that trust has an impact on the supervisor's intervention behavior, and the quality of the automation was raised as a factor in affecting  trust. 
Some early conceptual models of human trust in automation performing continuous process control include~\cite{lee1992trust,lee1994trust, muir1994trust} which related the various factors including the use and mis-use of automation, the trust of the operator in the automation, and the competence of the automation.  The authors in~\cite{muir1994trust} validated their conceptual model with two experiments, one qualitative and the other quantitative, involving an automatic process control~\cite{muir1996trust}.  The experiments showed that trust can be eroded by  temporary automation errors that have no consequence on the long term operation of the system, and it also showed that trust is not a binary variable but rather that there are multiple and possibly continuous levels of trust. The work in~\cite{lee1994trust} provided a mathematical model of how trust and confidence relate to the performance of the human-machine system. 
A contextually focused approach was taken in~\cite{cohen1998trust} which produced a relatively early computational model of trust. In~\cite{cohen1998trust}, trust is qualitatively  a ``context-specific'' argument against the performance of the system, and quantitatively a probability distribution of the correctness of the system actions given the system and the context or situation. In this framework, the problem of the acceptance of the output of decision support tools is not due to some steady state notion of under or over-trust but rather to a context-specific, transient notion of ``inappropriate trust'' in which there is a failure by the human operator to understand the conditions leading to either good or bad performance of the decision support tools.

\subsection{Failure Modes: Human Errors and Mode Confusion}  

The main approach to modeling human errors has been to use the phenotypical types as 
described by Hollnagel in~\cite{hollnagel1993phenotype}. 
The increase in reliability of aircraft systems has in turn 
highlighted cockpit crew error as a major cause of aviation incidents in recent years.   
A comprehensive framework that discusses human errors in aviation can be found in~\cite{wiegmann2001applying}.  The taxonomy framework, known as Human Factors Analysis and Classification System (HFACS), has its theoretical groundings in Reason's swiss-cheese model of human errors~\cite{reason1990human}, and groups human errors in aviation into 19 causal categories ranging from crew resource management errors~\cite{helmreich2000error} to individual errors in decision-making, perception and the execution of skill-based flying.  
Navigation errors are defined in~\cite{hooey2001post} as when the aircraft has moved into an uncleared taxiway or when the aircraft has deviated from the centerline in a cleared taxiway. As indicated by its definition, navigation errors occur only during surface or runway operations.   
Cockpit task management (CTM) errors occur when pilots initiate, monitor, prioritize and terminate concurrent cockpit tasks. The increase in automation of aircraft systems has resulted in a corresponding increase in CTM errors as a factor in aviation incidents. Empirical studies in~\cite{chou1996studies} show that CTM errors occurred in nearly 23\% of aviation incidents surveyed in the work. 

Mode confusion is the result of  
discrepancies between the human operator's mental model of the machine, and the actual state of the machine.  
Due to the increasing complexity of aviation systems,  mode confusion has been identified as a contributing 
factor to many recent aviation incidents~\cite{silva2015divergence}.  
Mode confusion was rigorously defined in~\cite{bredereke2002rigorous} using a failure refinement (or lack of) 
relation between two processes.  One process contains the safety-relevant  
system behavior as sensed by the human operator, 
and the latter process contains the operator's mental model of the safety-relevant system behavior. 
Leveson et al. in~\cite{leveson1997analyzing} discusses some of the common design flaws which 
can lead to mode confusion such as interface interpretation error (the 
user does not interpret the interface purpose correctly),  
indirect mode changes (automation mode changes that occur without explicit command from the user), 
inconsistent behaviors (automation exhibits the same behavior in many modes except for one or few),
and lack of appropriate feedback (automation does not provide the right visual/aural indications to the user that the mode has changed).

\section{Formal Models} 
\label{sec:formal-models}

This section samples humans models used in the \emph{formal verification} of human-machine systems.  

In formal verification, models of human, machine and the properties of interest (safety and/or performance) to be verified are captured using formal language(s) with unambiguous mathematical meaning. 
The human-machine system is verified when a  proof is generated that shows that the
system satisfies the properties of interest. This proof is typically either automatically generated using 
model-checking~\cite{clarke2018model}, or semi-automatically 
using a proof assistant~\cite{nipkow2002isabelle}.

\subsection{Formal Model of Tasks} 

We survey some efforts to formalize task and cognitive models of human operators.  
A more detailed review of various formal task and cognitive models can be found in this earlier survey paper~\cite{Bol13}.  
Formal models of human operators have been developed in the past using general languages such as process algebra~\cite{gunter2009specifying}, Petri-Nets~\cite{basnyat2007formal}, and more domain specific ones such as task analytic or cognitive modeling languages.  
This review focuses on the latter two as process algebra and Petri-Nets are already formal representations with well defined semantics and a variety of analysis tools already available to users, and the main issue is, therefore, the encoding process to represent a human model. 

Task analytic models are hierarchical representations of the observable set of human behaviors.  
In a task model, a high-level activity is decomposed into a set of composite actions 
which are further decomposed into a set of atomic actions. 
Some task analytic modeling languages are ConcurTaskTrees (CTT)~\cite{paterno1997concurtasktrees}, Operator Function Model (OFM)\cite{mitchell1986discrete}, Enhanced Operator Function Models (EOFM)~\cite{bolton2009enhanced}, and User Action Notation (UAN)~\cite{hix1993developing}.
Formal task analytic models can be encoded in a task modeling language, 
and then translated into a model checking or theorem proving representation for analysis and verification 
(see~\cite{palanque1995validating,ait2006formal,bolton2011systematic,ameur2003formal}). They can also be 
directly encoded in the formal language. 
Task model analyses in general are more limited by scalability issues than interface analyses 
since they cover the system beyond the interface. 
Nevertheless they have been extended to analyses with a team of operators or multiple operators (see~\cite{paterno1998formal,bass2011toward}) and errors~\cite{martinie2014fine}

\subsection{Formal Model of Cognition}

Lindsay and Connelly et al. in~\cite{cerone2005formal} applied model checking on a human operator model expressed in the cognitive architecture Operator Choice Model (OCM), and the approach was then applied in~\cite{lindsay2002modelling} to study erroneous behaviors of human air traffic controllers for various ATC tasks. OCM is not a generic cognitive architecture as it is specialized for ATC tasks.  Other works on more generic cognitive models such as Programmable User Models (PUM) include Blandford, Curzon, Butterworth et al. in~\cite{blandford2004models,butterworth1998role,butterworth2000demonstrating,rukvsenas2006formal}.  The verification in~\cite{butterworth1998role} was done using a theorem proving approach while~\cite{rukvsenas2006formal} used model-checking. One advantage that a cognitive model has over a task analytic model is that it can capture the organic underlying reasons for human errors rather than requiring researchers to introduce them explicitly into the model. The main drawback of cognitive models is that they are more complex hence less tractable by formal verification techniques. The work in~\cite{rukvsenas2009verification} is an example of  assessment of human errors in the context of ATM.
The work by Masci et al. in~\cite{masci2011modelling} uses the theorem-proving language PVS~\cite{owre:pvs:sri} to model a distributed cognitive model describing the information flow among multiple human operators. 
Distributed cognition~\cite{flor1991analyzing} is a conceptual framework in which cognition is a property of the whole system i.e. not just confined to the human minds. 
Additionally, a recent work in~\cite{Neo16} approached the problem by translating an existing cognitive model into a formal language representation.  The cognitive model, a SOAR model of air traffic contingency procedures performed by a human and automation team, was translated into a hybrid input-output automaton, which was then used to formally verify safety properties as the autonomy level of the system was increased.

\subsection{Formal Models of Workload} 

Formal models of workload is a relatively unexplored area.  
In~\cite{moore2014modeling}, the authors argued that one limitation of existing workload models 
is their level of resolution being too detailed and thus requiring
modeling efforts that are impractically time-consuming.  To address this limitation, 
they created an abstraction of a UAS-human teaming scenario with all of its actors modelled by finite-state machines augmented with workload metrics. Behavioral execution traces of the FSM model result in traces of workload metrics which can be monitored for violation of specification.  
Moore et. al however did not present any formal verification results. 
A more recent work~\cite{houser2018using} presented a novel approach which combined 
model-checking and simulation for the analysis of task loads and resource conflicts in 
air traffic control scenarios.  In this work, a trace of an air traffic simulation model  
for a particular scenario and time window is translated into a timed-automata model, which is used to model-check the entire ``neighborhood'' around the particular scenario for any violation of key properties. 
The counter-examples obtained by the model-checker are then fed back to the simulator for further analysis using higher fidelity models.

\subsection{Formal Model of Failure Modes}

Mode confusion have been analyzed formally using either model-checking and/or theorem proving (see~\cite{joshi2003mode,butler1998formal, campos2011modelling}). In all of those works, the human operator was not modelled 
modulo the human inputs to the machine. 
In~\cite{joshi2003mode} for example, model-checking and theorem proving were used to explore all the unexpected inputs that can trigger a particular flight guidance mode.  Other approaches towards formal analysis of mode confusion include modeling what the operator thinks is 
the state of the machine (also referred to as the mental model)~\cite{rushby2002using,degani2000pilot}, or how the operator accomplishes goals using the machine (also referred to as knowledge model)~\cite{javaux2002method}. Crow et al. in~\cite{crow2000models} proposed that mental models should be constructed from questionnaires to be incrementally refined and verified until the smallest safe mental model is found. However, extracting an accurate mental model from a human operator remains a challenge.   
In~\cite{combefis2015automatic}, the automation interface design tool ADEPT was used for the analysis of mode 
confusion.  Based on the mental model paradigm, the system and mental model are captured in ADEPT 
using an enriched labelled transition system called HMILTS. 

In the context of formal verification, one of the main approach to classifying and modeling human errors has been the binary phenotypes as described by Hollnagel in~\cite{hollnagel1993phenotype}. Several works~\cite{bastide2006error,bolton2008using,bolton2008formal} have incorporated Hollnagel's phenotypes of error into their task models and used  model-checking to determine whether the errors would lead to any violation of the specifications.

A more recent taxonomy of human errors developed to be more aligned with task analytic modeling and analysis 
techniques has been proposed in~\cite{bolton2017task}.  This taxonomy was first applied 
in~\cite{bolton2019formal} and then in~\cite{bolton2021formal}. In the latter work, reliability 
analysis and probabilistic model-checking were used to verify temporal-stochastic specifications such as 
the probability of an eventual failure being reached is less than some time bound.   
This supports the alternative view of human errors as probabilistic distributions of behavioral outputs under cognitive, motor and perceptual constraints.

\subsection{Formal Model of Knowledge} 

The epistemic approach~\cite{fagin2004reasoning} uses modal logic to capture and reason about the belief and knowledge of multi-agent systems. While epistemic modal logic has been studied for the past several decades~\cite{kripke1963semantical}, there appears to be little work which uses this framework to describe and reason about human behaviors in general. One possible reason is the lack of modeling tools that are intuitive to the domain expert.  This could be due to the lack of a grounded semantics for the systems under analysis which in our case is the human operator. The interpreted system~\cite{fagin2004reasoning} model has a well-known grounded semantics that has been used to model knowledge, belief and communications of computing systems.

Another possible reason is the dearth of efficient verification tools.  
Though recent model-checking literature~\cite{van2002model,kacprzak2004verification,wu2005model} have tackled the problem of verifying various language fragments with temporal and knowledge modalities, there are only a few modeling and verification tools that support epistemic and temporal modalities. One such tool is the MCMAS, described in~\cite{Lom15,Pen03}, which was developed for the verification of multi-agent interpreted systems. 

Perhaps another weakness of the epistemic approach is the problem of logical omniscience: an agent knows all the consequences of a set of logical assertions. Logical omniscience does not capture human characteristics such as bounded rationality, bounded reasoning and introspective forgetting.  Various modifications to the epistemic approach have been proposed to address  logical omniscience such as impossible possible worlds~\cite{hintikka1979impossible}, algorithmic knowledge~\cite{halpern1994algorithmic}, and non-classical propositional logic~\cite{fagin1995nonstandard}.  Other research works such as the one in~\cite{Van09} have proposed ways of enhancing the expressiveness of epistemic logic in order to succinctly capture introspective forgetting. 

Relative to task and workload models, the literature on modeling of 
knowledge and belief of human operators is very limited.  The work in~\cite{Pri14} extended CTLK (a temporal-epistemic first-order logic) with a notion of degree of
belief, and introduced a model-checking algorithm for such extended logic. 
The authors applied this extension to the Air France 447 (AF447) incident and duplicated
the NTSB conclusion that right before the crash, the pilots did
not believe any of the signals coming out of the cockpit display were true. 
The same case study was used to illustrate agent safety logic (ASL) 
in~\cite{ahrenbach2018formal}, a modal logic formalized in Coq~\cite{Coq:manual} 
with both epistemic and doxastic (belief) modalities combined 
with a safety logic based on the flight manual. 
ASL was used to illustrate the increasingly common phenomenon in aviation in which the pilot neither knows 
the state of the aircraft nor knows that he or she does not know the state of the aircraft. 

\subsection{Comparative summary}
\label{sec:formal-models-comparative-summary}
Table~\ref{tab:summary5} lists the models we have surveyed in this section. Each model is classified in terms of six key attributes that we believe are important for the formal verification of HMS. \emph{Focus} is the primary aspect of the human considered by the model. \emph{Formal verification} is the technique used to prove properties about the model, and can be either model-checking or theorem proving. By ``model-checking'' we mean any tool that performs symbolic or explicit state-space exploration, and that is typically fully automatic.  By ``theorem proving'', we meant a proof assistant tool such as PVS or Coq which is interactive. \emph{Timing} indicates whether the cited work has demonstrated the ability to verify non-trivial properties that involve time (either discrete or continuous). The use of temporal logic alone would not be considered as evidence that the model can handle properties about time. The same rationale applies 
to \emph{Probabilistic} and \emph{Epistemic}. The former indicates the ability to handle systematic uncertainty and answer queries about the probability of certain events. The latter indicates the ability to formally reason about the belief of the human. Finally, the \emph{Tool} column indicates whether the work has produced an integrated tool-chain with a frontend modeling environment that is connected to backend solver(s) for verification.

Table~\ref{tab:summary5} shows a general lack of support for uncertainty, both systematic and epistemic, which is expected given the inherent complexity of the verification problem for uncertain systems. Furthermore, there seem to be a lack of interest in the development of integrated tools bridging the gap between models and formal verification engines. 

\begin{table*}
    \centering 
    \caption{Summary of surveyed formal verification models}
\begin{tabular}{ |p{3cm}||c|c|c|c|c|c|}
    \hline
	Model & Focus & Formal Verification & Timing & Probabilistic & Epistemic &  Tool \\
	\hline
    Multiple~\cite{bolton2009enhanced,bolton2011systematic} & Task & Model checking & No & No & No & No \\
    \hline
    Multiple~\cite{palanque1995validating,ait2006formal,ameur2003formal} & Task & Theorem proving & No & No & No & No \\
	\hline
    Butterworth et al.~\cite{butterworth1998role} & Cognition & Theorem proving & Yes & No & No & No \\
    \hline
    Rukvsenas et al.~\cite{rukvsenas2006formal} & Cognition & Model checking & Yes & No & No & No \\
    \hline 
    Musci et al.~\cite{masci2011modelling} & Distributed Cognition & Theorem proving & No & No & Yes & No \\
    \hline 
    Multiple~\cite{rushby2002using,degani2000pilot} & Mode confusion & Model checking & No & No & No & No \\
    \hline 
    Houser et al.~\cite{houser2018using} & Workload & Model checking & Yes & No  & No & No \\
    \hline 
    Moore et al.~\cite{moore2014modeling} & Workload & No & Yes & No & No & No  \\
    \hline 
    Bolton, Zheng et al.~\cite{bolton2021formal} & Task + Error & Model checking & Yes & Yes  & No & No \\
    \hline 
    Multiple~\cite{bastide2006error,bolton2008using,bolton2008formal,bolton2019formal} & Task + Error & Model-checking & No & No & No & No \\
    \hline  
    Primero et al.~\cite{Pri14} & Beliefs & Model-checking & No & No & Yes & No \\
    \hline   
    Ahrenbach et al.~\cite{ahrenbach2018formal} & Beliefs & Theorem proving & No & No & Yes & No \\
    \hline 
    
\end{tabular}
\label{tab:summary5}
\end{table*}

\section{Simulation Models} 
\label{sec:simulations}
This section presents a survey of the simulation (or computational) 
models. By simulation, we also mean models that 
are compiled and then executed. 
Simulation models tend to contain more details than formal models, but 
verification using simulations is non-exhaustive in the sense that for 
most cases, it is not possible to simulate all possible inputs.  

\subsection{Models of Visual Processes}

Some of the more recent works in building visual attention models for simulation include a
cognitive architectural model with eye-tracking data in~\cite{Fle06}, a salience model~\cite{Itt98} for computerized scene analysis, and a Markov model 
in~\cite{Mel06}.  The salience model~\cite{Itt98} led to many further works in robotics
on duplicating the human visual capability.  The model is based on a biologically-plausible
architecture by Koch and Ullman~\cite{Koc87}. It assumes the existence of a master salience
map~\cite{Tho05} in
the posterior parietal cortex of primates that is influenced by top-down factors such goals and
knowledge.  This salience map is also influenced by a topographical map computed in a bottom-up
parallel process from feature maps (color, intensity, orientation, etc.) generated from the input
scenery. The regions on the salience map with the highest values correspond to a more rapid or
accurate response from the human operator. A recent significant advancement within the salience
framework is a fast computational model based only on the log spectrum of the input
image~\cite{Hou07}. 

Another important question that is addressed by perception 
modeling is \emph{visual attention allocation}, namely how much attention from the human operator a given area of the display captures relative to other areas. An analytic model for computing attention allocation is the
SEEV (Salience, Effort, Expectancy, and Value) model~\cite{Wic01}. This model was
developed for aviation applications, such as pilots in the flight deck~\cite{wickens2015noticing}, 
and has been extended to models of visual attention of
drivers~\cite{Mod06}.  The model is a simple function of four different variables: salience (the
likelihood that the visual event will attract attention), effort  (the amount of effort required
to re-allocate attention to the event), expectancy (the bandwidth and rate of information flow
generated by the event), and value (a measure of the cost of not processing the
event~\cite{Cas13}).  The
output of the function is the probability that a given area of the interface display will attract
the operator's attention.

Most recent developments in computer vision have been based on very large 
convolutional neural networks (CNN)~\cite{krizhevsky2012imagenet, he2016deep, he2017mask} trained on very large datasets~\cite{deng2009imagenet}
for tasks such as image classification and detection. The effective use of a black box model such as a CNN for the purpose of verifying 
human performance remains an open question.

\subsection{Models of Cognition}

For cognition, sophisticated simulation models such as 
cognitive architectures have been applied in
a wide variety of domains ranging from education~\cite{sweller1998cognitive} 
to aviation (see NASA and AFRL studies in~\cite{foyle2005human,gluck2006modeling}). 
Some examples of state-of-art cognitive architectures
include ACT-R~\cite{Leb98}, SOAR~\cite{Lai12,New92}, CHREST (Chunk Hierarchy and Retrieval
Structure)~\cite{Gob01}, ICARUS~\cite{Lan06}, and MIDAS~\cite{tyler1998midas}. 
The shared features of these cognitive architectures are:  
\begin{enumerate} 
\item A collection of underlying knowledge about the human intelligence such as how
short and long-term memories are formed and retrieved, and how the information from
memory gets processed.  
\item The production system representation which is an executable program with a set
of rules governing the behavior of an underlying phenomenon that it is describing.  
\item The problem space approach, 
	which includes a desired final state (also called goal), the initial state, 
	a set of the operators from a problem space 
	that when applied to a state lead to another state, and the search strategy over the problem space. 
\end{enumerate} 

Some cognitive architectures  are specialized for certain applications. The EPIC architecture of Meyer
and Kieras~\cite{Mey97}, for example, is especially suited for modeling multi-modal and
multi-task performance. Another example is the GOMS (Goals, Operators, Methods,
and Selection) architecture~\cite{Car83}, a simple cognitive architecture used to make qualitative
and quantitative predictions for normative or repetitive cognitive tasks in the
domain of human-computer interaction. In the context of aviation, the GOMS
framework has been used in the analysis of an advanced automated cockpit~\cite{Irv94},
and the ACT-R architecture has been used in modeling the differences in
skill acquisition of an ATC task~\cite{Taa02}. Cognitive architectures have also
incorporated other aspects of the human operator including, but not limited to,
perception, sensing, and actuation~\cite{And04}. Using a unified cognitive architecture
such as ACT-R, allows for developing a computational models of a human operator
executing a task in a particular domain.

\subsection{Models of Decision Making Processes}

Mathematical
representations of the lens model was developed in the works of Hammond and Hursch~\cite{Ham64}, and has been applied
to the study of air traffic control systems such as the work in~\cite{Bis03}. In this work, the authors quantify the
differences between the decisions made by a human air traffic controller against the decisions of the
automated air-collision alert algorithm. The mathematical model has the following structure: the
human judgment is assumed to be a probabilistic affine function of the cues, the environmental
criterion under judgment is assumed to be another probabilistic affine function of the cues, and the
measure of the accuracy of the human judgment is a nonlinear function of the correlation of the two
function outputs, the predictability of the two functions, and the correlation between the errors of
the two function outputs.

Another approach involves building expert systems, which are computer programs that mimic expert decision-making.  
In an expert system, the decision-making process of the human is captured by an inference process,
driven by if-then-else rules that use known facts from a knowledge database. 
In domains such as aviation, expert systems~\cite{endsley1987application} 
have been created to automate some of the functionality that were previously allocated to the human. 
In that sense, an expert system is a functional model of human decision-making. 

Simulation models of human decision-making 
have also utilized formalisms ranging from stochastic processes to continuous dynamical systems. 
One such modeling framework is the decision field theory (DFT)~\cite{busemeyer1993decision} which attempts to capture the dynamics of the human cognitive process in decision-making using a diffusion process, and has been used to predict response time and course of action under timing constraints and uncertainties.  In the context of air
traffic control (ATC) in the National Airspace System (NAS), models ranging from a
Partially Observable Markov-Decision Process (POMDP) in~\cite{Nie04}, to a Dynamic Bayesian Network
(DBN) in~\cite{Aco08} have been used. Furthermore, there are game-theoretic models such as the one in~\cite{Tom98}, and dynamical system models of human decision making such as the dead-reckoning navigational filter model used in~\cite{Pri01} or the Markov chain of Kalman filters used in~\cite{Pen99}.  Finally, a quantum dynamics model of human decision making (i.e. a model that describes the evolution of probabilistic amplitudes) was presented in~\cite{busemeyer2006quantum} and compared with a Markovian model. 

\subsection{Models of Tasks} 

Task analysis, and its specialized sub-field of cognitive task analysis (CTA)~\cite{schraagen2000cognitive} is a large collection of methods and tools for extracting expert knowledge, thought processes, and goals in the performance of tasks. In another words, it is an approach to describe the physical and cognitive tasks to accomplish an activity.   CTA is done for both observed behavior and latent cognitive functions.  It has been applied widely in the aviation domain~\cite{leiden2002information,seamster2017applied} as a knowledge capturing mechanism. 
With increased frequency, CTA is used by airlines and the  air traffic control community for the design of automation and training on the flight deck, operations, and the flight control tower.  For example, in~\cite{deutsch2005single}, the results of a CTA are used as inputs for modeling human operator performance in approach, landing, ATC, and ground operations of a single piloted aircraft. Task analysis are typically performed to capture expert knowledge such as task definitions and decomposition of missions.   Task definitions and decomposition are important steps to build task models for tools such as IMPRINT or other discrete-event simulators so that 
performance, typically workload, can be assessed for the scenario or application at hand. 

In~\cite{martinie2014fine}, a task hierarchical notation called HAMSTER augments the typical 
task hierarchy modeling formalism with information about system failures and human errors to analyze their effects on operator performance.

\subsection{Models of Workload}

There are numerous mental workload modeling methodology and tools, based on the multiple resources theory (MRT), for the simulation and prediction of operator performance.  They are especially useful for cases where it is not feasible to perform experiments on the actual system, environment and operators.  The task analysis workload (TAWL) framework ~\cite{hamilton1990task} and the TAWL operator simulation system (TOSS) was an early  methodology developed for the prediction of operator workload.  In the methodology, mission segments are decomposed into individual tasks, that are then modeled and simulated. This methodology bears resemblance to the task network modeling paradigm in the operator workload prediction tool IMPRINT~\cite{mitchell2000mental}.  Cognitive architectures such as the MIDAS performance tool~\cite{tyler1998midas} and ACT-R have also incorporated workload modeling methodology based on the MRT. 

Various computational models of multiple resource theory can be found in~\cite{sarno1995role,wickens2002multiple, horrey2003multiple} where~\cite{wickens2002multiple} is a simpler version of the model in~\cite{sarno1995role} and applied to the prediction of driving performance and interference in~\cite{horrey2003multiple}. These computational models can be distinguished from tools such as IMPRINT in that they can be used to predict more than just resource demand, but also task interference and re-allocation of resources.  Unlike most of the mental workload tools, MRT computational models can capture both cases when residual capacities exist in the human operator as well as when the operator is overloaded. The principle of multiple-resources has been used in various applications such as the development of cognitive architecture models for predicting workload~\cite{jo2012quantitative},  analysis of new cockpit display designs~\cite{wickens2003attentional}, and air traffic safety management~\cite{blom2001human}.

It was noted in~\cite{meshkati1988eclectic} that different performance metrics across different tasks make standardization of a performance-based workload measure difficult across domains.  Subjective evaluation techniques such as SWAT~\cite{reid1988subjective} or the NASA-TLX (task load index)~\cite{hart1988development} are methods that rely on self-reported ratings from operators, and/or observers. This is the most common technique used for the measurement of workload.  There are both one-dimensional and multi-dimensional workload rating scales in the literature. The latter typically aggregates the individual ratings into a single final workload value using either pair-wise weightings~\cite{hart1988development} or conjoint analysis~\cite{reid1988subjective}. The multi-dimensional scaled ratings are based on the assumption that the operator can evaluate the individual components of the workload better than the overall workload. Finally, the psycho-physiological techniques~\cite{dussault2005eeg,iqbal2004task,jorna1992spectral,may1990eye,vicente1987spectral} attempt to quantify workload by interpreting the measures of physical and neurological cues such as heart rate, EEG output, eye movements, pupil response, respiration rate, blood pressure, and sinus abnormalities.

\subsection{Models of Situational Awareness}
An early computational model of SA can be found in the work by Shively et al.~\cite{Shi97}.
This model is build within the MIDAS environment, and 
became the foundations for the more recent
computational model A-SA~\cite{McC02}. The A-SA consists of a low-level attention
module coupled with a high-level belief module. The attention and belief
modules are based on existing models of cognition and perception.  For example,
the attention module in A-SA is borrowed from Bundesen's 1990 work on the theory of
visual attention~\cite{Bun90} which has similarities with the more advanced SEEV model
introduced earlier.  The output of the model is a SA decay curve, in which the
rate of decay is a function of the distance between the evidences produced by
the visual module and the operator's current internal model of the world.  The
A-SA model has been mostly applied to applications in aviation such as pilot errors~\cite{Wic08}.  

\subsection{Models of Trust}


In~\cite{wiegmann2001automated,bisantz2001assessment,yeh2001display} the relations between the many variables in a human-machine system such as workload, difficulty of the task, system failure types, automation reliability, and interface display, are related to the notion of trust in the automation. 
A dynamical model was discussed in the survey paper~\cite{lee2004trust} where trust is dependent on a dynamical interaction between the operator, the interface, the automation, and the context. A closed-loop model of trust versus reliance was provided in which the interaction with the automation influences trust which in turn affects the interaction with the automation. Although there are many conceptual models of trust, only few computational models exist in the literature. In~\cite{gao2006extending}, a stochastic process model of decision-making in the context of human supervisory control, i.e. a binary decision process between manual control and automation control, was used to capture trust in a dynamic setting.  This model, which extends decision field theory to take into account prior decisions in a sequential decisions process, replicates several empirical nonlinear relations between trust, self-confidence and reliance. In~\cite{israelsen2021introducing}, a real-time multi-modal model of trust was produced based on structured equation modeling (SEM)~\cite{hoyle1995structural} which is a collection of statistical modeling and inference techniques that has had a long history of usage in the social sciences~\cite{morgan2015counterfactuals}.

\subsection{Integrated Modeling and Simulation Environments}

A recent trend in human modeling research is to focus on the integration of
theories, models, and tools together into a complete human performance modeling
environment. Cognitive architectures such as SOAR and ACT-R are some of the
most popular examples of integrated modeling environments. 
Another popular approach is based on a task network model where nodes (tasks) are events with triggering conditions, side effects during and post-execution, workload values, and probability of failures, and connections are transitions between the tasks. These approach is general and can readily accept many
theories, models and data from the human operator modeling literature. Workload assessment is done by running Monte-Carlo simulations on the task network model using a discrete-event simulator. The tool IMPRINT~\cite{Mit04} and its early
precursor SAINT~\cite{Pri74} are some examples of discrete-event modeling environment which 
have been specialized for assessing workload.  

Recent integration efforts connect distinct modeling and simulation environments together to handle more complex system-of-systems.  An argument for this approach is made in the introduction of~\cite{Leb05}.  As the system
being modeled becomes more complex, one modeling or simulation 
paradigm may not cover all the possible domains in the system.  Extending one paradigm to cover the new domains may not be possible since its capability might be stretched beyond
those for which it is a natural fit. The approach to address this problem is the
integration of other environments that are already well-suited for those
new domains.  An example of such effort is the IMPRINT/ACT-R integration done
in~\cite{Lau06} which was used to model and simulate an approach and landing problem provided by NASA. IMPRINT be naturally used for modeling all sequences of tasks at hand, while ACT-R is a natural environment for predicting human performance for each task. 
Another example is the composite cognitive
architecture QN-MHRP from~\cite{Liu06} which connects a queuing network model
with a cognitive architecture in order to study human performance on 
concurrent activities. 
Queuing networks are well-suited for modeling
parallel activities, while cognitive architectures are 
well-suited for predicting a person's action on a specific task. 

Other more specialized integrated environments include the MIDAS tool by NASA
Ames~\cite{Tyl98}, which is used in the design and analysis of cockpits and ATC
interfaces.  The distributed operator model architecture (D-OMAR), which is a
multi-tasking extension of the event-based modeling architecture OMAR~\cite{The96}, has
been used for modeling pilots in various studies in aviation, one of which
is the NASA human performance modeling study reported in~\cite{Lei05}.  This study was part of a
multi-year effort~\cite{Foy07}, which detailed the modeling
challenges and solutions of five teams, each using a different human performance
simulation and modeling environment (ACT-R, ACT-R/IMPRINT, Air MIDAS, D-OMAR and A-SA), to 
model and predict pilot performance and/or errors in two different contexts: 
instrumented approach and landing, and taxiway operations.
WMC or Work Models that Compute~\cite{pritchett2014modeling,pritchett2014work} 
is another integrated simulation framework that has been 
used to study human performance in aviation~\cite{lee2017simulating}. 
WMC allows the modeling of both the human and autonomous agents.  
The hybrid timing of its simulation engine enables simulation of both continuous-time dynamical systems and 
event-driven agents. 

\subsection{Applications to Human-Aviation Systems} 

An early study of single-pilot operations in the context of evaluating new advanced cockpit and control technologies for military helicopters can be found in~\cite{haworth1986investigation}.  In this study, test pilots were used in experimental simulations, and several subjective ratings scales including SWAT were used to evaluate cockpit workload. It was found that despite the advances in the cockpit, single-pilot operations increased workload. 
A model-based approach was taken in~\cite{deutsch2005single} to evaluate single-pilot commercial operations, where a two-pilots model built using the D-OMAR architecture was used as the baseline for evaluating performance in the single-pilot case and prior cognitive task analysis of approach, landing, and taxiing in~\cite{leiden2002information} was used as insight for the human performance models. 
It has been speculated in~\cite{harris2007human} that single-pilot commercial aircraft can be designed using mostly existing technology and available human factors models by  re-designing the flight deck to re-allocate some pilot functionality to automation and/or ground control. 

Navigation errors have been modeled in D-OMAR~\cite{deutsch2002modeling} and ACT-R~\cite{byrne2005using}.  In the former, navigation errors are captured in the model as the result of an unconscious winner-take-all selection of the wrong ``intention-to-act'', where each ``intention-to-act'' is some pre-generated cognitive plan to be executed. In the latter, the errors are revealed as the result of complex interactions between the cognitive model of the pilot, co-pilot, air traffic controller, and the model of the airport environment.

While experimental studies about single-pilot operations (SPO) in aircraft goes back decades~\cite{haworth1986investigation}, there are only a few works in the literature that have attempted the quantitative modeling and prediction of pilot performance in a single-pilot cockpit.  In both~\cite{stimpson2016assessing} and ~\cite{wolter2015validated}, a task analysis approach was taken.  
In~\cite{wolter2015validated}, task analysis of single-pilot concept of operations architectures was performed.  The scenarios  modeled in this work were related to the approach and landing portion of a flight into Denver airport. Task models were visualized in the environment Mirco Saint Sharp.  Performance measures based on  task counts and workload were compared against a baseline scenario of current day crew complement. 
In~\cite{stimpson2016assessing}, the modeling and simulation framework Pilot-Autonomy Workload Simulation (PAWS) measured the utilization of tasks, which has been applied in a prior work~\cite{cummings2007developing} to approximately measure a pilot's workload.  A custom discrete-event simulator was built to simulate the completions of tasks of various agents over the course of an entire flight from take-off to landing. The framework enabled a flexible configuration of different operational architectures, contingencies in scenarios, and degree of autonomy of the cockpit.  

The MDP framework was used in~\cite{feng2015controller} to model the interactions between a human operator and a UAV for the automatic synthesis of controllers. The human operator was modelled using a DTMC, and the UAV as MDP.  The two systems in closed-loop form a product MDP, which is then processed by a MDP synthesis algorithm to yield a controller for the UAV.  

Recent works in~\cite{boussemart2008behavioral,boussemart2011predictive} have also modelled human behaviors using the HMM approach.  In~\cite{boussemart2008behavioral}, a HMM was used to model the human supervisory control (HSC) of unmanned vehicles  and then used for predicting the future actions of the operator.  In~\cite{boussemart2011predictive} the modeling of HSC is extended using a hidden semi-Markov model in order to capture the timing of the transitions between the states which is important when the operational tempo of the HSC needs to be taken into account.

\subsection{Comparative summary}
Table \ref{tab:summary2} lists the models we have surveyed in this section. The attributes we have used to classify simulation models are similar to the ones in Section \ref{sec:formal-models-comparative-summary}, with the following differences. \emph{Formalism} refers to the structure of the computational model used in simulation. \emph{Stochastic} indicates whether the model can accommodate uncertainties in the inputs, outputs, and states of the system, and whether simulation supports them. Similarly,  \emph{Epistemic} refers to the ability to model and simulate  knowledge and beliefs. Finally, the \emph{Tool} column indicates whether  there is an integrated tool-chain with a frontend modeling environment that is connected to compilation, execution and simulation back-ends.

We observe a general good support for time and systematic uncertainty. This is expected given that simulation models often rely on discrete event simulation engines, that probabilistic events can be handled using random number generators, and that uncertainty quantification can therefore be done using Monte-Carlo methods. However, handling more complex epistemic properties seems to be out of reach. We also observe a pretty good availability of tools which are typically easier to build than in the case of formal verification. 

\begin{table*}
    \centering 
    \caption{Summary of surveyed simulation models}
\begin{tabular}{ |p{3cm}||p{3cm}|c|c|c|c|c|}
    \hline
	Model & Focus & Formalism & Timing & Stochastic & Epistemic & Tool \\
    \hline 
    Salience~\cite{Itt98} & Visual Attention & Neural Network & Yes & No & No & Yes \\ 
    \hline 
    SEEV~\cite{Wic01,wickens2009nt}& Visual Attention & Analytic & Yes & No & No & No \\
    \hline 
    CNNs~\cite{krizhevsky2012imagenet} & Vision & Neural Network & Yes & No & No & Yes \\ 
    \hline 
    ACT-R, SOAR, CHREST, ICARUS& General Purpose & Cognitive Arch. & Yes & Yes & No & Yes \\ 
	\hline
    EPIC, GOMS & Tasks & Cognitive Arch. & Yes & ? & No & Yes \\
    \hline 
    LENS~\cite{Bis03} & Decision & Multi-linear regression & No & No & No \\
    \hline 
    Multiple~\cite{Nie04,Aco08} & Decision & Stochastic process & Yes & Yes & No & No \\
    \hline 
    Tomlin et al.~\cite{Tom98} & Decision & Game-Theoretic & Yes & No & No & No \\
    \hline 
    Multiple~\cite{Pri01,Pen99}  & Decision & Dynamical Systems & Yes & No & No & No \\
    \hline 
    Busemeyer~\cite{busemeyer2006quantum}  & Decision & Quantum Dynamics & Yes & Yes & No & No \\
    \hline
    Shively et al.~\cite{Shi97} & SA & MIDAS & Yes & No & No & No \\
    \hline 
    A-SA~\cite{McC02,Wic08} & SA & Analytic & Yes & No & No & No \\
    \hline 
    Multiple~\cite{dussault2005eeg,iqbal2004task,jorna1992spectral,may1990eye,vicente1987spectral} & Workload & Empirical & Yes & No & No \\
    \hline 
    SWAT and NASA TLX & Workload & Surveyed Data & No & No & No & No \\
    \hline 
    TAWL and TOSS~\cite{hamilton1990task} & Workload & Task Network & Yes & No & No & No \\
    \hline 
    Gao et al.~\cite{gao2006extending} & Trust & Stochastic process & Yes & Yes & No & No \\
    \hline
    Israelsen et al.~\cite{israelsen2021introducing} & Trust & SEM & Yes & No & No & No \\
    \hline 
    Martinie et al.~\cite{martinie2014fine} & Tasks & Task Hierarchy & No & No & No & Yes \\
    \hline 
    Task Analysis & Tasks & Task Hierarchy & No & No & No & Yes \\
	\hline
    PAWS~\cite{stimpson2016assessing} & Single Pilot Workload & Discrete-Event & Yes & ? & No & Yes \\
    \hline
    OMAR, IMPRINT, SAINT & General Task + Workload & Discrete-Event & Yes &  Yes & No & Yes \\
    \hline
    D-OMAR & Distributed Pilot Task + Workload & Discrete-Event & Yes &  Yes & No & Yes \\
     \hline
    Feng et al.~\cite{feng2015controller} & Mission Execution Time  & Stochastic process & Yes &  Yes & No & No \\
    \hline 
    QN-MHRP from~\cite{Liu06} & Concurrent Tasks & Integrated & Yes & Yes & No & Yes \\ 
    \hline 
    MIDAS & Cockpit and ATC Interface & Integrated & Yes & Yes & No & Yes \\
    \hline
    ACT-R/IMPRINT & General Purpose & Integrated & Yes & Yes & No & Yes \\
	\hline 
    WMC~\cite{pritchett2014work} & General Purpose & Integrated & Yes & ? & No & Yes \\
    \hline
\end{tabular}
\label{tab:summary2}
\end{table*} 



\section{Conclusions and Recommendations}
\label{sec:conclusion}
\label{sec:modeling-of-human-autonomy-systems}
We provided an overview of human modeling including fundamental theories of human cognition, decision-making, perception and actuation, as well as the state-of-art human operator modeling and human factors techniques for assessing human performance in aviation, and for the design, verification and validation of human-machine systems. 

We have found that human operator for human-\emph{automation} systems is much more common than for human-\emph{autonomy} systems. By automation, we mean systems helping humans to perform a limited set of tasks in a limited set of environments. In this case, engineers can provide dedicate software programs to handle each case. When the number of goals and environments makes this standard approach impractical, then the system must include perception and decision-making, and we refer to it as autonomy. In this latter case, there is a need to reason about what the system and the human ``know'' about the environment, what the system ``believes'' that the human ``knows'', what the human believes that the system knows, and so on. Also, we have found that, while general purpose human models exist (such as cognitive architectural models),  they tend to be difficult to construct as they may require a considerable knowledge capturing effort, they are rather brittle, and are not suitable for systematic design, verification and validation of human-machine systems.

System-theoretic models, such as stochastic processes, are easier to construct. However, they may not scale well to industrial-scale applications and generalize poorly with respect to the task and the context. Formal methods models have been primarily built for the analysis and verification of interactions between human and existing automation. There is a lack of formal modeling of human operator for assessing  performance and capabilities in scenarios involving the operation of autonomous systems including single-pilot applications. Furthermore, the approaches we have surveyed tend to analyze the combined human-machine system at one level of abstraction, hitting the well-known state explosion problem.

Based on the outcomes of this survey, we feel that further research and development efforts are required to deliver a unified framework for modeling and verification of human-autonomy systems. We make the following recommendations  

\noindent
\emph{Declarative and compositional specifications.} 
Declarative modeling~\cite{kowalski1979algorithm,fahland2009declarative} enables us to provide a description of the capabilities of the human operator rather than details behind the mechanism which implements those capabilities. The expected increase in the complexity of human-autonomy systems makes testing and simulation less effective due to the large possible set of environments to be considered. Declarative models enable \emph{Formal methods}~\cite{Peled} are exhaustive and lead to strong guarantees on safety and/or performance at the expense of computational complexity. Tractability problems can be tackled with appropriate abstractions and compositional frameworks such as contract-based design (CBD) \cite{benveniste2018contracts}.  
One difficulty facing compositional modeling of human operators is that workload measures of tasks are difficult to compose. Typically, getting the workload of two or more tasks being performed concurrently requires re-collecting workload ratings for the combination of tasks being performed. Multiple-resource models provide a somewhat limited computational way of composing workloads. However, those compositional techniques are still experimental and have not been extensively validated. 

\noindent
\emph{New specification languages to account for different kinds of uncertainty and time.} 
As machine exhibits more human-like traits such as using knowledge to make 
complex decisions, there emerge a need to 
reason about not only about what the system knows, but also what the human operator knows about the system, and what the system knows about what the human operator knows about the system, and so on. 
Modeling knowledge and higher order of belief can be achieved using the framework provided by epistemic logic \cite{fagin2004reasoning}. This framework has been used, for instance, in \cite{fisher2013verifying}. Models of humans should be able to describe the state of knowledge which includes what the human knows about the knowledge (not the vivid facts) of the machine. There are several tools that have appeared in the area of verification of epistemic dynamic systems (see for example \cite{vcermak2014mcmas}) that could be used to perform formal verification. However, the problem is in modeling the epistemic state in a declarative way. Moreover, issues such as awareness \cite{fagin1987belief,sim1997epistemic} has to be taken into account, and modeling techniques and verification tools for modeling awareness seem to be lacking. Furthermore, the declarative modeling language should mix epistemic modalities with several others including temporal and probabilistic ones, leading to verification problems that become intractable. Research in both areas of modeling and decision procedures is needed to enable formal verification of these new class of models. 

\noindent
\emph{Human-factors, subject-matter experts, and engineering interfaces.} 
The analysis of human-machine systems will require collaboration among several experts in different areas such as human-factors (HF), subject-matter (SM), and engineers (E). An integrated modeling and verification environment will need to provide appropriate interfaces to all. HF and SM experts may not have the appropriate background to build models amenable to formal analysis. Similarly, they may not be able to interpret results generated by formal verification tools. 

Interpretability of analysis results by HF and SM experts will be essential given the current state of research in human modeling as outlined in this survey. There is a certain degree of model uncertainty that may lead to incorrect analysis results. Thus, we believe that it is important to provide the right level of explanation to experts in various field to assess the results and identify potential modeling errors.  

\bibliographystyle{IEEEtran}
\bibliography{refs/tim,refs/human_operator,refs/fm}

\end{document}